\documentclass[prb,twocolumn,draft,amsmath,showpacs]{revtex4}
\usepackage{graphics}

\begin{document}

\bibliographystyle{prsty}
\input epsf

\title {Temperature dependence and anisotropy of the bulk upper critical field $H_{c2}$ of MgB$_2$}

\author {A.V. Sologubenko, J. Jun, S.M. Kazakov, J. Karpinski, H.R. 
Ott}
\affiliation{Laboratorium f\"ur Festk\"orperphysik, ETH H\"onggerberg,
CH-8093 Z\"urich, Switzerland}

\date{\today}

\begin{abstract}
The bulk upper critical field $H_{c2}(T)$ of superconducting  MgB$_2$ and 
its anisotropy are established by 
analyzing experimental data on the 
temperature and magnetic field dependences of 
the $ab$-plane thermal conductivity of a single-crystalline sample
in external magnetic fields oriented both parallel ($H_{c2}^{c}$) and perpendicular 
($H_{c2}^{ab}$) to 
the $c$ axis of the hexagonal lattice. 
From numerical fits we deduce the anisotropy 
ratio $\gamma_{0}=H_{c2}^{ab}(0)/ H_{c2}^{c}(0)=4.2$ at $T=0$~K.
Both the values and the temperature dependences of $H^{c}_{c2}$ and $H^{ab}_{c2}$  are
distinctly different from previous claims based on measurements of the 
electrical resistivity.
\end{abstract}
\maketitle

Since the recent discovery of superconductivity of MgB$_2$ with a critical
temperature $T_c\simeq 40$~K,\cite{Nagamatsu01} a large number of 
experimental results on different properties of this compound has been reported in the literature. 
Most experiments were made using powder or polycrystalline samples.
The hexagonal crystal structure 
of MgB$_{2}$, however, is expected to  cause pronounced anisotropies 
in the 
electrical and 
magnetic properties, which can unambiguously be probed only by 
experiments using single crystals. In particular,  
the upper critical field $H_{c2}(T)$ is 
an  important parameter for characterizing 
the superconducting state of type II superconductors.\cite{Shulga01cm,Manske01cm,Haas02}
For anisotropic materials, such as hexagonal  MgB$_{2}$, 
the values of $H_{c2}$ may vary considerably for different 
orientations  of the external magnetic field $H$. 
Choosing the field 
directions either perpendicular  or parallel to the $c$-axis, the 
anisotropy may be expressed by a parameter $\gamma = 
H_{c2}^{ab}/H_{c2}^{c}$, which, in the most general case, may be 
temperature dependent.
Earlier experimental results, mainly based on measurements of the 
electrical resistivity $\rho(T,H)$, 
have resulted in a broad range of values 
of $\gamma$ and of extrapolated zero-temperature values of  
$H^{c}_{c2}(0)$ and $H^{ab}_{c2}(0)$ (for a review, 
see Ref.~\onlinecite{Buzea01}).
Most of these experiments, also on single crystals,\cite{Lee01,Xu01,Kim01cm} indicate a 
positive curvature of $H_{c2}(T)$ in a wide range of temperature  below $T_{c}$ 
and correspondingly, rather high critical fields at $T=0$. 
Attempts to explain these features have lead to theoretical work 
suggesting the existence of 
some soft bosonic modes\cite{Shulga01cm} and even unconventional 
mechanisms of superconductivity have been considered.\cite{Zhao01cm}

In this paper we present an evaluation of $H_{c2}^{c}(T)$  and 
$H_{c2}^{ab}(T)$ of single crystalline MgB$_{2}$,  based on measurements 
of the thermal conductivity $\kappa(H,T)$.  
Complementary results of $\rho(T,H)$, obtained on  the same 
single-crystalline sample, indicate that electrical transport measurements 
are not well suited to 
probe the bulk upper critical field $H_{c2}(T)$ of MgB$_{2}$. 
Inspecting the temperature dependences of  both
$H_{c2}^{c}(T)$ and $H_{c2}^{ab}(T)$ close to to $T_{c}$, our results indicate that even the 
zero-field critical temperature $T_{c}(0)$ of the bulk may be lower 
than commonly believed up to now. This  indicates that, in relation 
with superconductivity of MgB$_{2}$, surface effects must be 
considered.

The thermal conductivity was measured in the basal plane of hexagonal MgB$_2$ 
exposed to varying magnetic fields $H$, oriented parallel and perpendicular 
to the basal $ab$-plane with a small misalignments of $3.5 \pm 0.5^{\circ}$
between the field directions and the orientation of the plane.
A standard uniaxial heat flow method, as described in Ref.~\onlinecite{Gianno00}, 
was used for the $\kappa(H,T)$ measurements. The temperature 
difference between the two thermometers was about 1\% of the absolute average 
temperature.  
The measurements of the electrical resistivity $\rho(H,T)$ were 
made using a 4-contact scheme and a $dc$-current of 
density 50~ A/cm$^2$ in the $ab$-plane with $H$ along the $c$-direction.
The investigated single
crystal has lateral dimensions of $0.5\times 0.17\times 0.035$ mm$^{3}$
and was grown employing a high-pressure cubic anvil technique as
described elsewhere.\cite{Karpinski01}

\begin{figure}[t]
 \begin{center}
  \leavevmode
  \epsfxsize=1\columnwidth \epsfbox {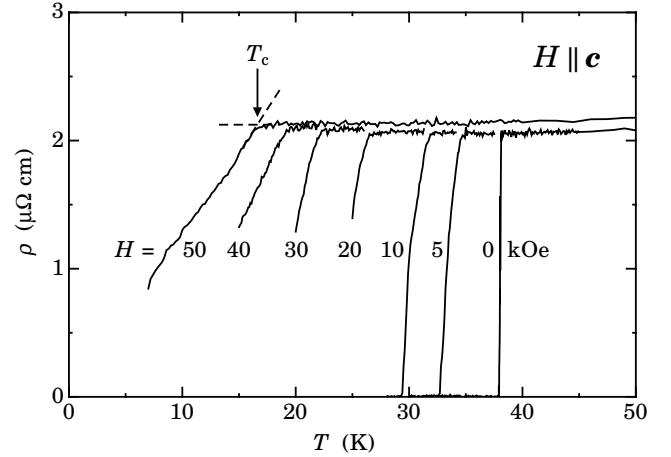}
   \caption{
   The low temperature 
  electrical resistivity of MgB$_{2}$ and its magnetic field 
  dependence for a current in the $ab$-plane. 
  The broken lines for $H=50$~kOe indicate how the onset of the 
  resistive transition defining $H^{c}_{\rho}$ has been established.
  }
\label{R}
\end{center}
\end{figure}

\begin{figure}[t]
 \begin{center}
  \leavevmode
  \epsfxsize=1\columnwidth \epsfbox {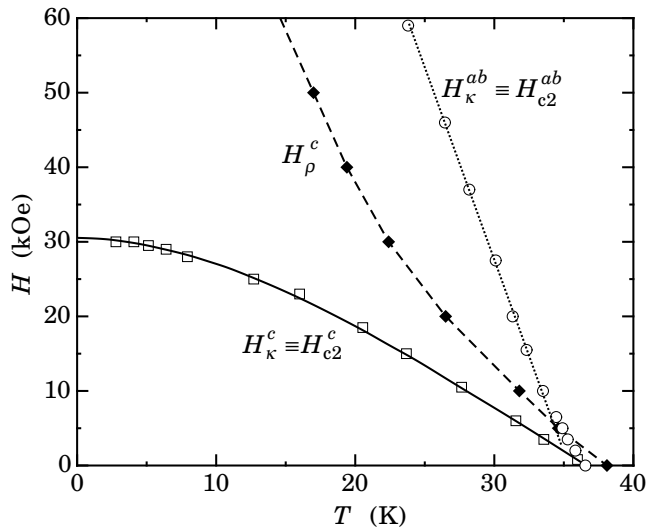}
   \caption{
  Critical fields $H_{\rho}$ and $H_{\kappa} \equiv H_{c2}$, as determined
  from  electrical resistivity and thermal conductivity measurements, respectively.
  The solid line 
  is compatible with calculations due to Helfand and Werthamer\cite{Helfand66}.
  The dashed an dotted lines are to guide the eye.
  }
\label{Hc2}
\end{center}
\end{figure}

Low-temperature $\rho(T)$ curves measured in
constant external magnetic fields $H$  are presented in Fig.~\ref{R} 
for $T < 50$~K.
The zero-field resistive superconducting transition  at $T_c=38.1$~K is
rather narrow ($\Delta T_{c} \sim 0.15$~K), but the  application of magnetic 
fields broadens the transition considerably. 
If the field dependence of the onset of the resistive transition,
as illustrated in Fig.~\ref{R} for $H=5$~T, is plotted in an [$H,T$] 
diagram, the curve denoted as $H^{c}_{\rho}$ in Fig.~\ref{Hc2} is 
obtained. These $H^{c}_{\rho}$ data are qualitatively and 
quantitatively very similar to results previously obtained 
on single crystals,\cite{Lee01,Xu01,Kim01cm} in particular with respect to 
the positive curvature of  $H^{c}_{\rho}(T)$. In these earlier works 
$H^{c}_{\rho}(T)$ was associated with $H^{c}_{c2}(T)$.

\begin{figure}[t]
 \begin{center}
  \leavevmode
  \epsfxsize=1\columnwidth \epsfbox {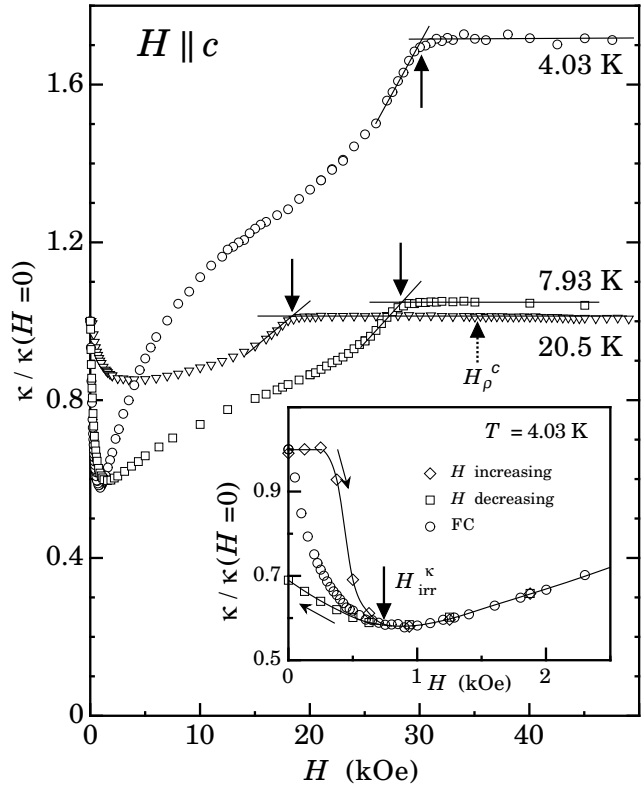}
   \caption{
    The magnetic field dependence of the thermal conductivity 
    $\kappa(H)$ for $H\parallel c$ at $T= 4.03$, 7.93  and 
  20.5~K.
  The solid vertical arrows mark $H^{c}_{\kappa}$, set equal to the 
  upper critical field
  $H_{c2}^{c}$.
   The dotted vertical arrow denotes $H^{c}_{\rho}$ (see
   Fig.~\ref{R}).
   The inset demonstrates the irreversible behavior
  of $\kappa(H)$ below $H_{\rm irr}^{\kappa}$  for $T=$4.03~K.
  }
\label{KH_c}
\end{center}
\end{figure}

\begin{figure}[t]
 \begin{center}
  \leavevmode
  \epsfxsize=1\columnwidth \epsfbox {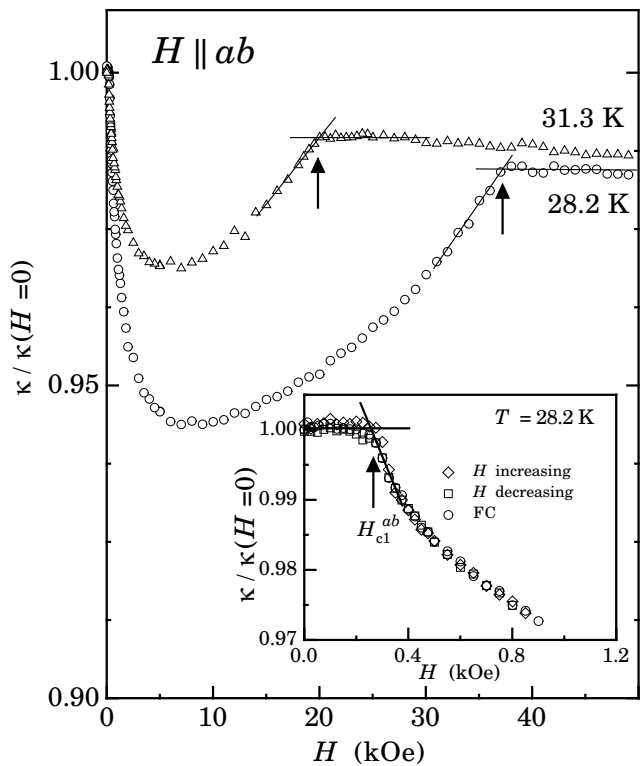}
   \caption{
  The magnetic field dependence of the thermal conductivity 
    $\kappa(H)$ for $H \perp c$ at $T= 28.2$ and 31.3~K.
  The solid vertical arrows mark $H^{ab}_{\kappa}$, set equal to the 
  upper critical field
  $H_{c2}^{ab}$.
   The inset demonstrates the small hysteresis 
  of $\kappa(H)$ at low fields for $T=28.2$~K.
  }
\label{KH_ab}
\end{center}
\end{figure}

The $H$-dependence of the thermal conductivity was measured at selected constant 
temperatures in the range between 2 and 50~K and in fields up to 60~kOe. 
Representative $\kappa(H)$ curves at selected 
temperatures are displayed in Figs.~\ref{KH_c} and \ref{KH_ab} for $H 
\parallel c$ and $H \parallel ab$, respectively. 
As demonstrated in the inset of Fig.~\ref{KH_c}, 
a hysteretic behavior of $\kappa(H)$, caused by vortex pinning, 
is observed in the low-field regime for $H \parallel c$.
In order to avoid ambiguities, each new field setting at a constant 
temperature was achieved by heating the sample to the normal state 
above 50~K, and  subsequently cooling it to the set temperature in the 
chosen field.
In this way, 
a smooth variation of $\kappa(H)$, as demonstrated 
by the open circles (FC) in the inset of Fig.~\ref{KH_c}, was obtained.
The field values 
$H^{\kappa}_{\rm irr}$, below which the irreversibility is discernible, are 
rather low. For $T=4.03$~K, e.g., $H^{\kappa}_{\rm irr} \sim 
0.7$~kOe. 
At elevated temperatures and for $H \parallel ab$ the 
irreversibilities are reduced,  
as demonstrated in the inset of 
Fig.~\ref{KH_ab}.

The curves presented in Figs.~\ref{KH_c} and \ref{KH_ab}
reveal the general features observed at all temperatures below 
38.1~K.  
Starting at $H=0$,
$\kappa$ drops 
with a steep slope and, after passing through a minimum, 
increases again until a region of very weak 
field dependence above some critical field, denoted as 
$H_{\kappa}$ is reached. It is remarkable that, for each 
temperature, $H^{c}_{\kappa}$ 
is distinctly  lower than  $H^{c}_{\rho}$ and that no distinct feature 
of $\kappa(H)$ is observed in the region of $H^{c}_{\rho}$. 
This is explicitly demonstrated in Fig.~\ref{KH_c}.
With increasing temperature,  $H_{\kappa}$ decreases towards
zero as $T$ approaches $T_{c}$.  
This general $\kappa(H)$ features are 
typical for type II superconductors and 
can be explained as follows. \cite{Vinen71} 
The thermal conductivity of a 
superconductor is due to itinerant electrons ($\kappa_{e}$) 
and phonons ($\kappa_{\rm ph}$). 
Enhancing  $H$ from 
zero eventually causes the formation of vortices in the bulk of a type 
II superconductor. After zero-field cooling, the first vortices form at the lower critical field $H_{c1}$. 
Consequentially, some  
additional scattering of phonons by normal electrons in the cores of 
the vortices will reduce $\kappa_{\rm ph}$. 
With further increasing field the decrease of $\kappa_{\rm ph}$ is 
compensated by an enhancement of $\kappa_{e}$. 
Above $ H_{c2}$, in the normal state, the field dependence of both 
$\kappa_{\rm ph}$ and $\kappa_{e}$ is expected to be weak. 
The overall behavior of the $\kappa(H)$ curves shown in 
Figs.~\ref{KH_c} and \ref{KH_ab} reflects these expectations, and as 
may be seen, $\kappa(H)$ is virtually field independent for $H > 
H_{\kappa}$.

A more complete analysis of the $\kappa(H)$ data will be 
presented in a forthcoming paper.\cite{Sologubenko01_c} 
Here, we concentrate on the opportunity that  
these data allow for a reliable 
evaluation of the bulk upper critical field $H_{c2}(T)$, which obviously 
coincides with $H_{\kappa}(T)$ as derived from our $\kappa(H)$ curves 
for both field orientations. 
It may be seen that  $H^{c}_{c2}(T)\equiv H^{c}_{\kappa}(T)$ is distinctly 
different from $H^{c}_{\rho}(T)$. 
The solid line in Fig.~\ref{Hc2}, representing a general prediction 
for $H_{c2}(T)$ of a conventional type II superconductor in the case 
where the coherence length $\xi$ and  the electron mean free path 
$\ell$ are of similar magnitude,\cite{Helfand66}
is 
in fair agreement with the measured  $H_{c2}^{c}(T)$.
It is obvious that
$H^{c}_{\rho}(T)$ does not follow the same general $T$-dependence. 
Since thermal conductivity experiments probe the bulk of the sample,  
it is  $H^{c}_{\kappa}(T)$ rather than   
$H^{c}_{\rho}(T)$ that  ought to be identified as  
the upper critical field $H^{c}_{c2}(T)$. 
Recent magnetization measurements\cite{Angst01cm} on single-crystals 
of MgB$_{2}$ using a torque 
magnetometer result in values and a temperature dependence of 
$H^{c}_{c2}(T)$ that are consistent with our 
$H^{c}_{\kappa}(T)$ and thus support our conclusion.
Employing the equation given by anisotropic 
Ginzburg-Landau theory 
$H_{c2}(\theta) = \left[ (\sin \theta /H^{c}_{c2} )^2  + (\cos \theta 
/H^{ab}_{c2} )^2     \right]^{-1/2}$, where $\theta$ is the angle between 
the magnetic field and the $ab$-plane,\cite{TinkhamBook} we estimate 
the errors in calculating $H_{c2}$ caused by the above mentioned misalignment of 
$3.5 \pm 0.5^{\circ}$ to be about $3\pm 1\%$ for 
$H^{ab}_{c2}$ and below 0.2\% for $H^{c}_{c2}$.  

As displayed in Fig.~\ref{Hc2},  
at temperatures above about 27~K, $H_{c2}^{c}$ varies linearly with 
temperature with a slope 
$dH_{c2}^{c}/dT = -1.17$~kOe/K. 
This behavior leads to an extrapolated 
zero-field $T'_c=36.6$~K,  1.5~K below $T_{c}$ obtained from 
$\rho(T,0)$.
Using the equations which are  given by the  Ginzburg-Landau theory 
considering anisotropies, \cite{TinkhamBook}
$\xi_{ab}(T)  = (\Phi_{0}/2\pi H_{c2}^{c}(T))^{1/2}$ and 
$ \xi_{ab}(T) = 0.74 (1-T/T_c)^{-1/2} \xi_{ab, 0}$, where $\xi_{ab}$ is the coherence length in the 
basal $ab$-plane, 
we obtain the zero-temperature value of
$\xi_{ab, 0}=11.8$~nm.

Turning to the temperature dependence of the critical field  
$H_{c2}^{ab}(T)$ for $H \perp c$, 
we again note a sizable temperature interval where   
$H_{c2}^{ab}(T)$ varies linearly with $T$. This is  emphasized by the dotted line in 
Fig.~\ref{Hc2}. 
The slope $dH_{c2}^{ab}/dT=-5.15$~kOe/K gives
$\sqrt{\xi_{ab, 0}\xi_{c, 0}}= 5.75$~nm and therefore, 
the zero-temperature value of the $c$-axis correlation length 
$\xi_{c, 0}=2.8$~nm. Another important parameter which can be 
estimated from our $\kappa(H)$ data is the lower critical field $H^{ab}_{c1}$, as 
demonstrated in the inset of Fig.~\ref{KH_ab}. In the temperature 
region between 28 and 35~K, where an evaluation of $H^{ab}_{c1}$ with 
reasonable accuracy of about $\pm 10\%$  was possible,  
$H^{ab}_{c2}/H^{ab}_{c1}\approx 130$. From this 
ratio, using the equation  
$H_{c2}/H_{c1}=2\kappa_{\rm GL}^{2}/\ln{\kappa_{\rm GL}}$,\cite{TinkhamBook}
the parameter $\kappa^{ab}_{\rm GL}$ of the Ginzburg-Landau 
theory is estimated to be about 13.

As may be seen in Fig.~\ref{Hc2}, above  approximately 33~K,  
$H_{c2}^{ab}(T)$ deviates from the linear in $T$ variation and, with 
increasing temperature,  approaches zero also at 
$T'_{c}$ defined above.
This is reflected in the temperature dependence of the anisotropy 
ratio $\gamma$, which seems to decrease with T  
approaching $T'_{c}$.
The positive curvature of $H_{c2}(T)$ is typical for strongly 
anisotropic, layered 
superconductors\cite{Woollam74} and has often been  explained 
in terms of the  Lawrence-Doniach model,\cite{Lawrence70} which treats a layered 
superconductor as a stacked array of weakly coupled two-dimensional 
superconducting sheets. Various other theoretical models have been proposed 
to explain this feature (for a critical review see, e.g., Ref.~\onlinecite{Brandow98}).
At this point we cannot commit ourselves to any of these models.
It is important, however, that the anomaly is absent for $H \parallel c$
and small and restricted to a rather narrow temperature region for $H 
\perp c$. At lower temperatures, with decreasing temperature
the anisotropy ratio $\gamma(T)$ tends to a constant value and 
is approaching 
$\gamma_{0}= H^{ab}_{c2}(0)/H^{c}_{c2}(0) = 
\xi_{ab, 0}/\xi_{c, 0} = 4.2$.

The extrapolation to zero-temperature gives  $H_{c2}^{c}(0) \approx 
31$~kOe, a considerably lower value 
than is typically claimed  for MgB$_{2}$.\cite{Buzea01} Exceptions 
are the reports of Refs.~\onlinecite{Budko01Hc2,Simon01}. 
The  rather low value of $H^{c}_{c2}(0)$ 
and the observation of a  Helfand-Werthamer-type\cite{Helfand66} 
temperature dependence of $H^{c}_{c2}(T)$
have important consequences for 
possible models  of the superconducting state of MgB$_{2}$. 

Our result obviously questions the intrinsic nature of 
$H_{c2}(T)$ derived from measurements of $\rho(T)$.
Since the resistive transition is not manifest
in $\kappa(T)$, which may be considered as a bulk property, $H_{\rho}(T)$ must
correspond to a minor fraction of an additional phase (or phases) with enhanced $H_{c2}$ 
and $T_{c}$. The spatial extension of this phase is, however,  large enough to 
short-circuit the electrical current path and produce a narrow superconducting 
transition at a temperature $T_{c}$ higher than the bulk transition 
temperature $T'_{c}$. Based on an analysis of their magnetization 
and $ac$-susceptibility data on polycrystalline samples, the authors of 
Ref.~\onlinecite{Fuchs01} came 
to a similar conclusion.  The most likely origin of  the second phase 
with enhanced superconducting parameters seems to be related 
to surface effects. A considerable enhancement of the electronic
density of states near the Fermi level and, therefore, an enhanced 
trend to 
superconductivity at the surface of MgB$_{2}$ have been predicted\cite{Kim01,Bascones01,Silkin01}, 
in agreement with our observations.

In conclusion, we observe a striking disagreement in the 
values and the temperature dependences of the upper critical field 
$H^{c}_{c2}$ of MgB$_{2}$ 
evaluated from results of electrical and thermal conductivity 
measurements on the same sample. 
The shape of  $H_{c2}^{c}(T)$ as established by $\kappa(H)$ with $H 
\parallel c$ does not 
reveal an anomalous  positive curvature near $T_{c}$ 
and therefore no  exotic mechanism needs to be involved to explain the 
upper critical field, at least not for the bulk. Our data also 
indicate that the bulk transition temperature $T'_{c}$ is lower than 
$T_{c}$ obtained from results of $\rho(T)$ measurements.

\acknowledgments
We acknowledge useful discussions with I.L. Landau, M. Angst and R. 
Monnier.
This work was financially supported in part by
the Schweizerische Nationalfonds zur F\"orderung der Wissenschaftlichen
Forschung.


\end{document}